\documentclass[a4paper,12pt]{article}
\usepackage{graphicx,amsmath,lineno,color}
\usepackage{authblk}
\usepackage{xspace}
\usepackage{subfig}
\usepackage{upgreek}
\usepackage[hidelinks,colorlinks=true,linkcolor=blue,urlcolor=blue,citecolor=blue]{hyperref}
\usepackage[textheight=25 cm, textwidth=18cm, headheight=50pt, headsep=10pt, footskip=32pt]{geometry}\hypersetup{colorlinks=true}

\newcommand*{\betastar}{\ensuremath{\beta^{\star}}\xspace}

\newcommand*{\sigmatot}{\ensuremath{\sigma_{\textrm{tot}}}\xspace}
\newcommand*{\sigmael}{\ensuremath{\sigma_{\textrm{el}}}\xspace}
\newcommand*{\sigmainel}{\ensuremath{\sigma_{\textrm{inel}}}\xspace}
\newcommand*{\dsigmadt}{\ensuremath{{\textrm{d}}\sigma/\textrm{{d}}t}\xspace}
\newcommand*{\electronvolt}{\text{e\kern-0.1em V}}
\newcommand*{\TeV}{\ensuremath{\text{T\electronvolt}}}
\newcommand*{\GeV}{\ensuremath{\text{G\electronvolt}}}

\title{Determination of the total cross section and $\rho$-parameter from elastic scattering in $pp$ collisions at $\sqrt{s}=13$ TeV  with the ATLAS detector}
\author{Hasko Stenzel, on behalf of the ATLAS Collaboration}
\affil{II. Physikalisches Institut, Justus-Liebig-Universit\"at Giessen\\
  Heinrich-Buff-Ring 16, D-35392 Giessen, Germany\\
  email \href{mailto:Hasko.Stenzel@cern.ch}{Hasko.Stenzel@cern.ch}}

\begin{document}
\maketitle
  
\abstract{A new measurement of elastic $pp$ scattering at $\sqrt{s} = 13$ TeV with the ATLAS-ALFA detector is presented. The measurement was performed using data recorded in a special run of the LHC with $\beta^\star = 2.5$ km. 
The elastic cross-section was measured differentially in the Mandelstam $t$ variable and from a fit to 
${\textrm{d}}\sigma/\textrm{d}t$ the total cross section, the $\rho$-parameter and parameters of the nuclear slope are determined.
The results for $\sigma_{\textrm{tot}}$ and $\rho$ are
\begin{equation*}
\sigma_{\textrm{tot}}(pp\rightarrow X) =  \mbox{104.7} \; \pm 1.1 \; \mbox{mb} , 
 \;  \; \rho =  \mbox{0.098} \; \pm 0.011 . 
\end{equation*}
The energy evolution of $\sigma_{\textrm{tot}}$ and $\rho$, connected through 
dispersion relations, is compared to several models. Furthermore, the total inelastic 
cross section is determined from the difference of the total and elastic cross section, and 
the ratio of the elastic to total cross section is calculated.
}

\section{Introduction}
This proceedings contribution presents a new measurement by ATLAS~\cite{ATLAS} using $pp$ collision data at $\sqrt{s}= 13$~\TeV, corresponding to an integrated luminosity of 340~$\upmu$b$^{-1}$. The elastic data were recorded 
by the ALFA~\cite{ALFA} subdetector of ATLAS in a special run using a beam optics with $\beta^\star = 2.5$ km.
In elastic scattering the distribution of the four-momentum transfer $t$, defined by $-t = \left(\theta^\star \times p \right)^2$ 
where $\theta^\star$ is the scattering angle and $p$ the beam momentum, is measured.    
This beam optics allows to extend the reach in small values of $|t|$ to such small scattering angles that the Coulomb interaction starts to play a role. The coverage of the Coulomb--nuclear interference (CNI) region enables the extraction 
of the $\rho$ parameter defined as 
\begin{equation}
 \rho = \left. \frac{\textrm{Re}[f_{\textrm{el}}(t)]}{\textrm{Im}[f_{\textrm{el}}(t)]}\right|_{t\rightarrow 0} ,\label{eq:rho}
\end{equation}
\noindent where $f_{\textrm{el}}(t)$ is the elastic-scattering amplitude.
The optical theorem connects the hadronic component of the total cross section $\sigmatot$ to the imaginary part of the scattering amplitude in the forward direction,
\begin{equation}
\sigmatot=4\pi \, \textrm{Im} \, [ f_{\textrm{el}}\left(t)]\right|_{t\rightarrow 0} .
\label{eq:OpticalTheorem}      
\end{equation}
In order to extract $\sigmatot$ and $\rho$ from a fit to the differential elastic cross section an ansatz on 
the shape of the scattering amplitude needs to be made. The present analysis uses for the nuclear amplitude an exponential function 
incorporating parameters to describe a $t$-dependent nuclear slope, 
which is found to describe well the curvature of the data observed at large $|t|$~\cite{ALFA13TeV2p5}.

\section{Experimental conditions}
The ALFA subdetector~\cite{ALFA} of ATLAS is a Roman Pot detector system installed in the LHC tunnel 
at a distance of about 240 m on either side of the interaction point. Two detector stations on each side house each two vertically moveable scintillating fibre detectors to measure trajectories of elastically scattered protons. For the data taking period with $\beta^\star = 2.5$ km the distance 
of the detectors to the beam orbit was about 1 mm. The data were recorded in four fills of the LHC. 
During each fill the beam was repeatedly scraped with primary collimators to remove the background 
from accidental halo protons from the sample of elastic candidates. 

\section{Data analysis}
The data were selected following stringent data quality requirements for events triggered by a 
left--right coincidence in the elastic back-to-back configuration with tracks reconstructed in four detectors forming a spectrometer arm~\cite{ALFA13TeV2p5}. Further geometrical requirements 
were applied to the left-right acollinearity. The irreducible background composed of events from 
central diffraction was estimated from simulation and background from accidental halo coincidences 
was determined in a data-driven method using templates. The total background level is below $0.1\%$. 
The detector alignment and reconstruction efficiency were also derived from data-driven methods. The reconstruction of the $t$-value requires the transport matrix elements of the beam optics, which was 
tuned in a procedure resulting in an effective beam optics using a fit to variables measured in ALFA. 

\section{Theoretical prediction}
The theoretical form of the $t$-dependence of the cross section is obtained by:
\begin{eqnarray}\label{eq:tgen}
\frac{\mathrm{d}\sigma}{\mathrm{d}t} & = & \frac{4\pi\alpha^2(\hbar c)^2}{| t |^2} \times G^4(t) \\ \nonumber
 & - &  \; \; \sigmatot \times \frac{\alpha G^2(t)}{|t|}\left[\sin\left(\alpha\phi(t)\right) + \rho \cos\left(\alpha\phi(t)\right) \right] \times \mathrm{e}^{\frac{-B|t| - Ct^{2} - D|t|^{3}}{2}} \\ \nonumber
 & + &  \; \; \sigmatot^2 \frac{1+\rho^2}{16\pi(\hbar c)^2} \times \mathrm{e}^{\left({-B|t|-Ct^{2}-D|t|^{3}}\right)}  ,
\end{eqnarray}
where $G$ is the proton form factor and $\phi$ the Coulomb phase. 
The first term in Eq.~\eqref{eq:tgen} corresponds to the Coulomb interaction, the second to the Coulomb--nuclear interference, 
and the last to the hadronic interaction. This parameterization is used to fit the differential elastic cross 
section to extract the physics parameters $\sigmatot$ and $\rho$, and the terms $B$, $C$ and $D$ relevant to the nuclear slope.

\section{Results}
\begin{figure}[htb!]
  \centering
\subfloat[Fit of $\dsigmadt$.]{
\includegraphics[width=0.49\textwidth]{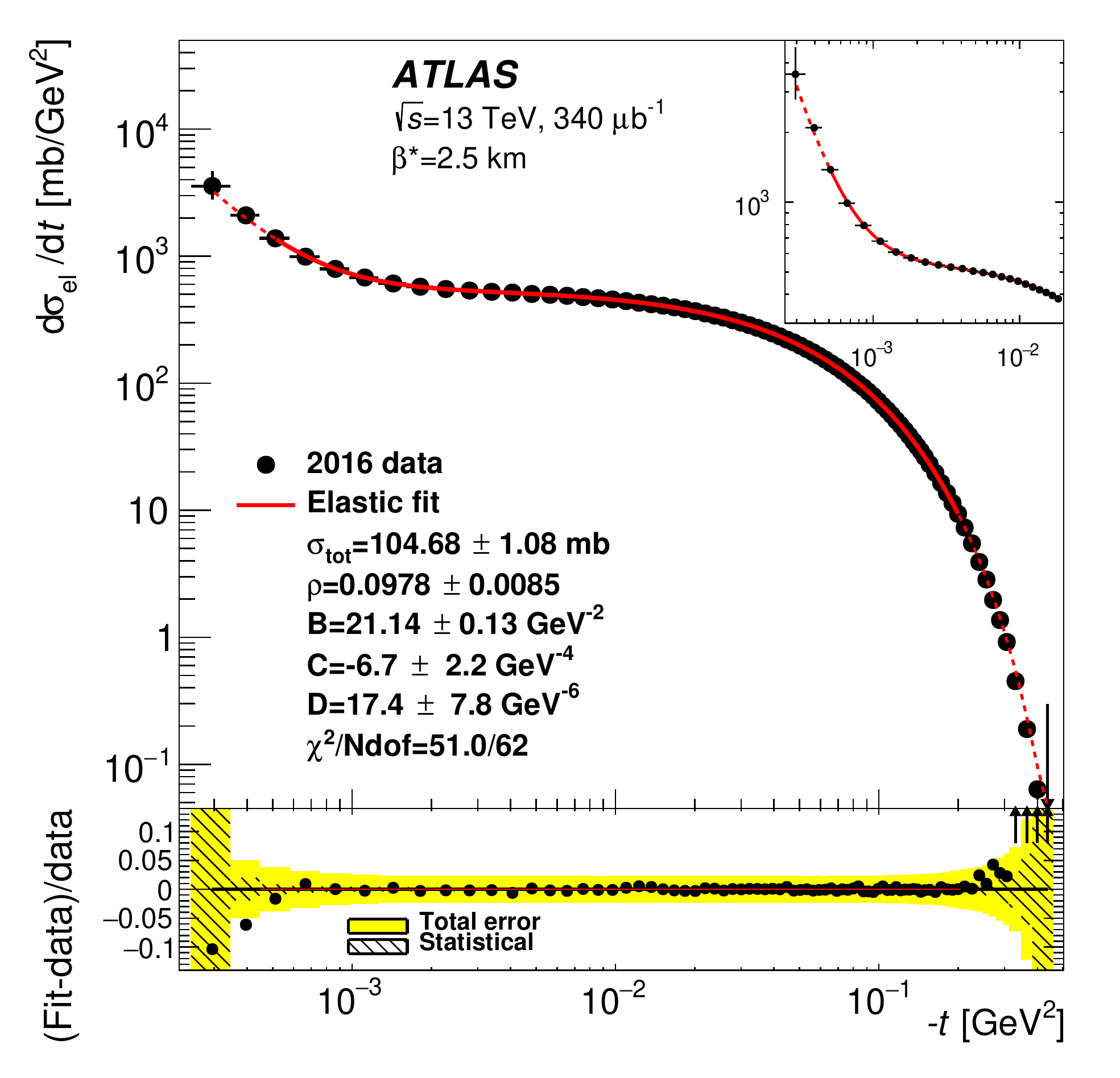}
\label{fig:tfit}
}
\subfloat[Relative difference to the reference]{
\includegraphics[width=0.49\textwidth]{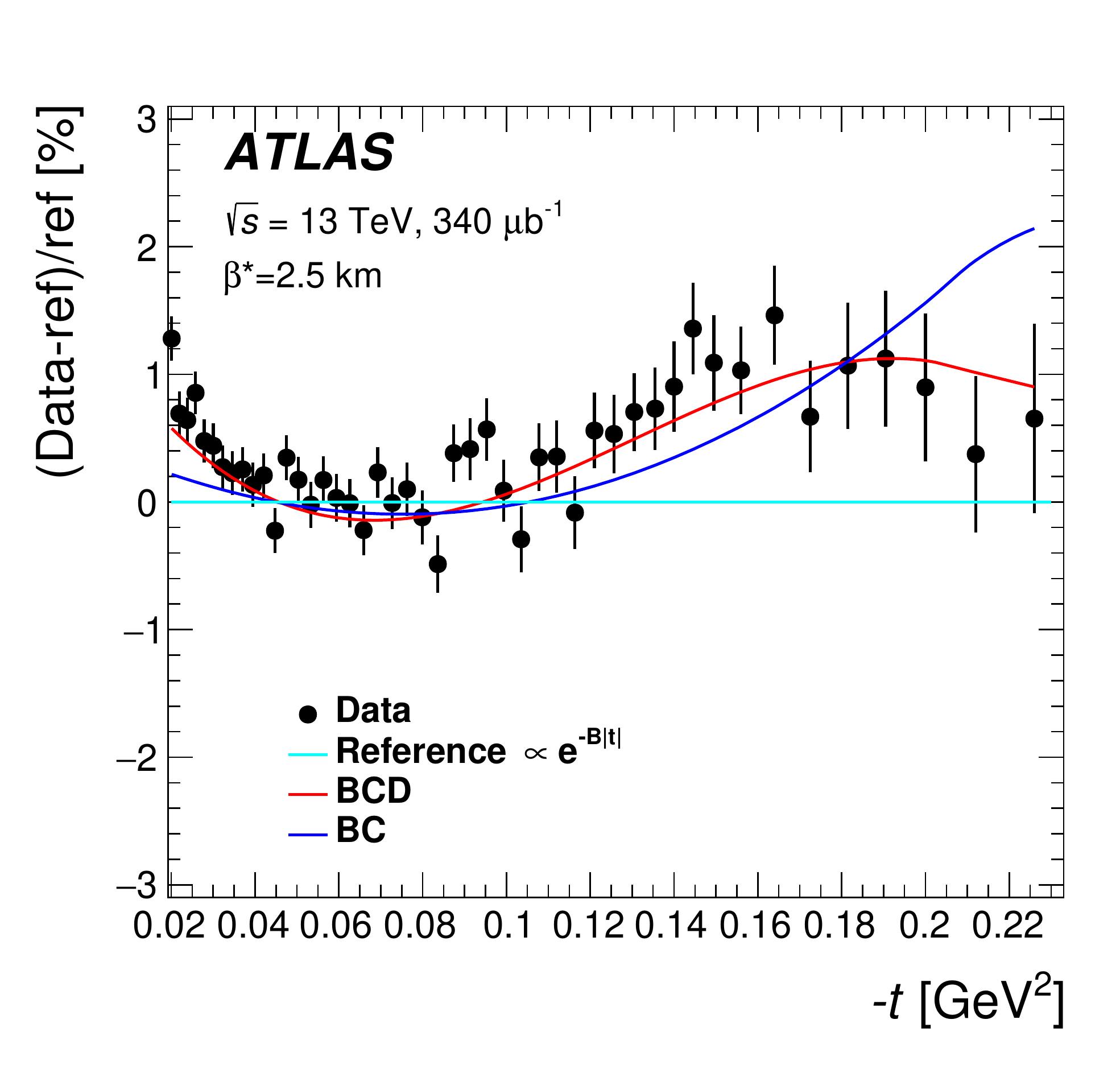}
\label{fig:curvature}
}
  \caption{\protect\subref{fig:tfit} A fit of the theoretical prediction with $\sigmatot$, $\rho$, $B$, $C$ and $D$ as free parameters to the
 differential elastic cross section.  
 \protect\subref{fig:curvature} The data normalized to a reference exponential function compared to models of the nuclear amplitude with a $t$-dependent 
  exponential slope~\cite{ALFA13TeV2p5}.} 
  \label{fig:tfit_nuisance_subtraction}
\end{figure}

The differential elastic cross section is calculated from the raw $t$-spectrum applying 
acceptance and unfolding corrections from simulation, correcting for the reconstruction 
and trigger efficiencies and by normalizing to the recorded integrated luminosity. The latter was determined with a precision of $2.15\%$ in a dedicated analysis, adapting techniques developed for high-luminosity running~\cite{lumi13TeV}. The experimental systematic uncertainty 
for $\dsigmadt$ accounts for the detector alignment, beam optics, reconstruction efficiency, 
beam emittance and energy, simulation modelling, detector resolution, background subtraction and luminosity. The main uncertainties are related to the luminosity and alignment. 
The physics parameters are extracted from the data by a profile fit using Eq.~\eqref{eq:tgen} which includes both  
statistical and experimental systematic uncertainties and their correlations. For each source of systematic uncertainty a nuisance parameter is fitted simultaneously with the physics parameters. The differential elastic cross section and fit of the theoretical prediction 
is shown in Figure~\ref{fig:tfit}. 
The fit range is chosen from $t = -4.5\cdot10^{-4}~\GeV^2$ to $t = -0.205~\GeV^2$ corresponding 
to the range with an acceptance above $10\%$. Theoretical uncertainties are evaluated by 
changing the assumptions on the form of the nuclear amplitude and by taking into account the 
uncertainties of the proton form factor and Coulomb phase parameterization. Also different 
models of the nuclear phase were considered. The results are summarized in Table~\ref{tab:profile_fit}.

An evidence for the $t$-dependent slope of $\dsigmadt$ motivating the introduction of the terms $B,~C$ and $D$ 
in the nuclear amplitude parametrization (see Eq.~\eqref{eq:tgen}) is shown in Figure~\ref{fig:curvature}. The 
data and models are normalized to a reference exponential function fitted in a restricted range at small $|t|$, 
where the slope is approximately constant. The precision of the present data require  
two additional parameters $C$ and $D$ to fit the observed curvature.
\begin{table}[tp]
  \begin{center}
  \caption{Results of the profile fit to the differential elastic cross section.}
    \begin{tabular}{l|ccccc}
      \hline 
   & $\sigmatot~[\mbox{mb}]$ & $\rho$ & $ B~[\GeV^{-2}]$ & $C~[\GeV^{-4}]$ & $D~[\GeV^{-6}]$\\ 
      \hline
 Central value  & 104.68~~~~ & 0.0978 & 21.14~~ & $-6.7$~~\, & 17.4~~ \\ \hline 
 Statistical error    & 0.22 & 0.0043 & 0.07 & 1.1 & 3.8 \\
 Experimental error    & 1.06 & 0.0073 & 0.11 & 1.9 & 6.8 \\
 Theoretical error      & 0.12 & 0.0064 & 0.01 & ~~0.04 & ~~0.15 \\ \hline
 Total error    & 1.09 & 0.0106 & 0.13 & 2.3 & 7.8 \\ \hline
      \end{tabular}
  \label{tab:profile_fit}
  \end{center}
\end{table}

The hadronic part of the total elastic cross section is extrapolated to the full phase space by integrating 
the nuclear part of the fitted prediction. The value of $\sigmael=27.27 \pm 1.14$ mb can be subtracted from 
the value of $\sigmatot$ to obtain the total inelastic cross section, which is found to be $\sigmainel = 77.41 \pm 1.09$ mb. From these measurements the ratio of the elastic to total cross section is calculated: 
$\sigmael/\sigmatot = 0.257 \pm 0.012$. 

\section{Interpretation}
\begin{figure}[h!]
  \centering
\subfloat[$\sigmatot$]{
\includegraphics[width=0.49\textwidth]{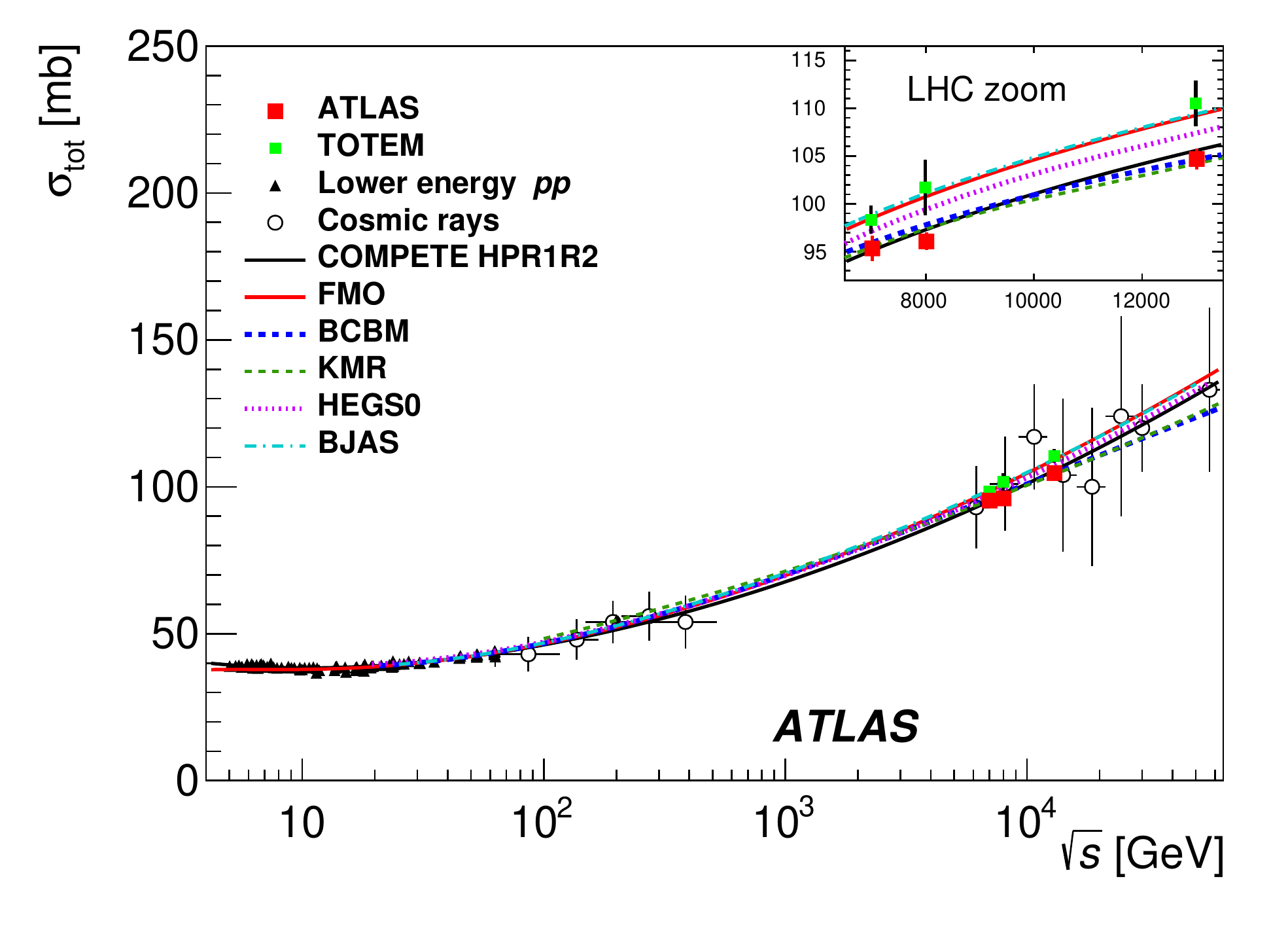}
\label{fig:stot}
}
\subfloat[$\rho$]{
\includegraphics[width=0.49\textwidth]{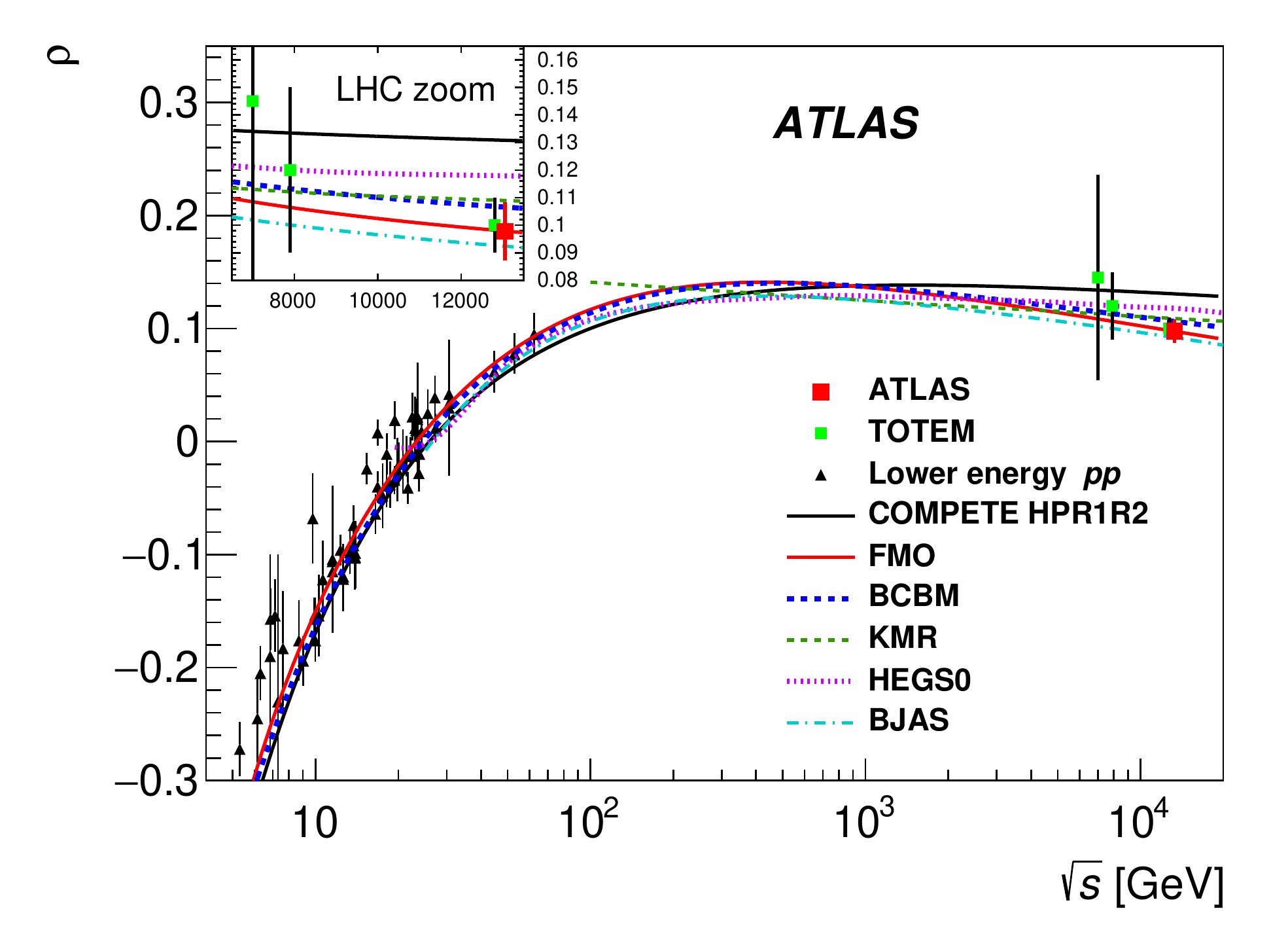}
\label{fig:rho}
}
\caption{Energy evolution of \protect\subref{fig:stot} the total cross section 
  and \protect\subref{fig:rho} the $\rho$-parameter compared to different model predictions~\cite{ALFA13TeV2p5}.} 
  \label{fig:running_models}
\end{figure}

The energy evolution of $\sigmatot$ and $\rho$ need to be studied simultaneously, because they 
are connected through dispersion relations, which are based on the fundamental principles of analyticity, 
crossing symmetry and unitarity of the elastic-scattering amplitude. Several models were developed on these principles to describe the energy evolution. The predictions of some widely used models are shown in 
Figure~\ref{fig:running_models}.    
The COMPETE collaboration~\cite{COMPETE} performed global fits to elastic-scattering data based on Regge theory using a 
crossing-even amplitude, leading to an energy evolution of $\sigmatot$ described at high energy by a combination of terms in $\ln s$ and $\ln^2 s$. The ALFA measurements of $\sigmatot$ are in good agreement with the COMPETE 
prediction, but it exceeds the $\rho$ value at $13~\TeV$ by about $3 \sigma$. The new value of the total cross section is about~5.8 mb lower than the measurement from the TOTEM Collaboration~\cite{TOTEM13TeV2p5}, 
corresponding approximately to a tension of $2.2\sigma$. 
The low value of $\rho$ can possibly be explained by a contribution from a crossing-odd amplitude in the scattering process, corresponding to the exchange of the Odderon. The FMO model~\cite{FMO} features a maximal Odderon exchange and was tuned to the TOTEM data up to $13~\TeV$, but it exceeds the ALFA $\sigmatot$ data. Another model proposed in Refs.~\cite{Block_and_Cahn,Bourrely_and_Martin}, derived in the context of rise of the total cross section observed at the ISR, using only a crossing-even amplitude, suggested the possibility of a damped amplitude of the form $\ln^2s/(1+\alpha\ln^2s)$, where $\alpha$ is a damping factor. A global fit to data shown in Figure~\ref{fig:running_models} was performed; it yields the best agreement with the data among the tested models. Quantitative results on the consistency of these 
and further models with the data are given in Ref.~\cite{ALFA13TeV2p5}. 

\section{Conclusion}
This contribution presents measurements of the total cross section, $\rho$-parameter and nuclear slope parameters 
using elastic $pp$ scattering data at $\sqrt{s}=13~\TeV$ recorded by the ALFA subdetector of ATLAS 
in 2016 in a special LHC run with $\betastar=2.5$~km optics, corresponding to an integrated luminosity of $340~\upmu$b$^{-1}$. 
From a fit to the differential elastic cross section in the range from $-t = 4.5 \cdot 10^{-4}~\GeV^{2}$ to $-t = 0.2~\GeV^{2}$,  
the total cross section and $\rho$-parameter are determined to be:
\begin{eqnarray*}
\sigmatot(pp\rightarrow X) & = &  \mbox{104.68} \; \pm \mbox{1.08} \; (\mbox{exp.})  \pm \mbox{0.12} \; (\mbox{th.})  \; \mbox{mb}, \\
\rho & = & \mbox{0.0978} \; \pm \mbox{0.0085} \; (\mbox{exp.})  \pm \mbox{0.0064} \; (\mbox{th.}) ,
\end{eqnarray*}
where the first error accounts for all experimental systematic uncertainties and includes the statistical component, and the second  
is related to the model uncertainties. 
This analysis included a dedicated luminosity determination for the $\betastar=2.5$~km run, which is directly used 
to normalize the cross-section measurements. A study of the form of the $t$-spectrum revealed the need to introduce a $t$-dependent parameterization  
of the exponential nuclear slope. It is found in this analysis that with two more parameters $C$ and $D$ in addition to $B$ used in previous measurements 
a satisfactory description is achieved; they are measured to be:
\begin{eqnarray*}
B & = &  \mbox{21.14} \; \pm \mbox{0.13}~\GeV^{-2}  , \\
C & = &  \mbox{$-6.7$} \; \pm \mbox{2.2}~\GeV^{-4}  , \\
D & = &  \mbox{17.4} \; \pm \mbox{7.8}~\GeV^{-6}  .
\end{eqnarray*}

The new data for $\sigmatot$ and $\rho$ are compared with lower-energy data, and the energy evolution of these data is analysed 
in the context of model studies of the evolution. 
This study shows that the commonly accepted energy evolution is in tension with 
the $13~\TeV$ elastic-scattering data. 
Further research is needed to understand the dynamics of strong interactions leading to the unexpectedly low value of $\rho$.

\end{document}